\documentclass[twocolumn,epsf,prb,amsmath,amssymb,showpacs]{revtex4}
\usepackage{amsmath}
\usepackage{amsfonts}
\usepackage{amsmath}
\usepackage{amssymb}
\usepackage{graphicx}
\usepackage{epsfig}
\usepackage{dcolumn}
\usepackage{rotate}
\usepackage{lscape}

\usepackage{bm}
\begin{document}
\title{Aharonov-Bohm oscillations in a mesoscopic ring with asymmetric arm-dependent injection}
\author{P. Vasilopoulos$^{1,2}$, O. K\'alm\'an$^{2,3}$,  F. M. Peeters$^2$, and M.
G. Benedict$^3$
\ \\
\ \\}
\affiliation
{$^{1}$Department of Physics, Concordia University,
1455 de Maisonneuve
Ouest, Montr\'eal$,$ Quebec Canada H3G 1M8 \\
$^2$Departement Fysica, Universiteit Antwerpen (Campus
Groenenborger)  Groenenborgerlaan 171 B-2020 Antwerpen, Belgium \\
$^3$Department of Theoretical Physics, University of
Szeged,
Tisza Lajos k\"or\'ut 84 H-6720 Szeged, Hungary}

\begin{abstract}
Electron transport through mesoscopic, one-dimensional rings with
asymmetric injection into the arms of the ring
is studied,  in the presence of
a Aharonov-Bohm flux, by means of an appropriate {\bf S} matrix.
This matrix is expressed in terms of two parameters one of which
($\lambda$) accounts phenomenologically for
this asymmetric injection 
into the arms of the ring. In addition, the effect
of a scatterer placed in one arm of the ring is  considered.
Explicit expressions are obtained for the transmission as  a
function of the incident electron energy, the magnetic field,  the
asymmetry parameter $\lambda$, and
 the strength of the scatterer. Results of the literature for
 symmetric rings are described by $\lambda=1$ and readily recovered.  We relate our results
 to rings of finite width.
\end{abstract}
\maketitle

\section{Introduction}

The  study of rings goes back to Aharonov and Bohm (AB) who
demonstrated the importance of vector potentials in quantum
mechanics \cite{aha}. This goes under the name AB effect and one
frequently uses the term AB oscillations for the  oscillations in
the resistance of a ring as a function of the flux penetrating the
interior of the ring. Another major study is now referred to as the
Aharonov-Casher\cite{aha1} effect which is similar to the AB effect
but it is due to the spin-orbit interaction (SOI). Other
developments concern the Berry phase \cite{ber}. With the
development of new fabrication techniques and the size reduction of
samples, rings are now very intensely studied especially in
connection with the SOI, see Refs. \onlinecite{fru} - \onlinecite{cho}  and
references cited therein.

In a quantum ring of finite width the connection between the
current-carrying leads and the ring can be complicated and may lead
to reflection at the lead-ring junction and to asymmetric injection
in the two arms of the ring. This asymmetry can be a consequence of the difference in length
between the upper and lower arms
or of fabrication defects but it can also be induced by  a magnetic field
as a consequence of the Lorentz force. Such an asymmetry was
demonstrated recently from a pure numerical treatment of the
transmission through a finite-width lead-ring system; it led to
incomplete AB oscillations due to partially destructive
interferences \cite{sza}.

To our knowledge all previous works that study transmission through a
ring employ an
${\bf S}$ matrix that is {\it symmetric} with respect to both arms
of the ring \cite{but}-\cite{col}. Though this may not be as
restrictive as it sounds,
it is more realistic to reexamine the problem using an ${\bf S}$
matrix that is {\it not symmetric} with respect to both arms of the
ring and possibly make a connection with the asymmetry mentioned
above \cite{sza}.

An asymmetry can be introduced by placing, e.g., one scatterer in
one arm of the ring or by locally applying a gate that affects the properties of  one arm. The scatterer may introduce important phase
shifts in the electron wave function and change drastically the
position and/or amplitude of the AB oscillations.
This has been thoroughly investigated theoretically
\cite{col} and experimentally \cite{ya}. Although we will consider such a case,
here we are mainly concerned  with {\it asymmetric} current injection into the arms of the ring through one of the leads, cf. Fig. 1.

In view  of the above, we propose a 1D model in which asymmetries
due to fabrication,
scatterers, and especially asymmetric
current injection through one of the leads are parametrized by a
small number of parameters but which leads to explicit analytical
results. This has the advantage of being more useful to
experimentalists than the pure numerical treatment  of Ref.
\onlinecite{sza}.

 In the next section  we formulate the problem and derive analytic
 expressions for the transmission amplitude.  We present analytical results
 in Sec. III and numerical results in Secs. IV and  V.
 Concluding remarks follow in Sec. VI.

\section{Formulation of the problem}

At each junction of a lead with the ring, indicated by the triangles
in Fig. 1, we have three
outgoing waves with amplitudes
$(\alpha^\prime, \beta^\prime, \gamma^\prime)=\bm{ \alpha}^\prime$
and three
 incoming waves with amplitudes $(\alpha, \beta, \gamma)=
\bm{ \alpha}$. They are related by a $3\times 3$ ${\bf S}$ matrix in
the manner
\begin{figure}[tpb]
\includegraphics*[width=80mm]{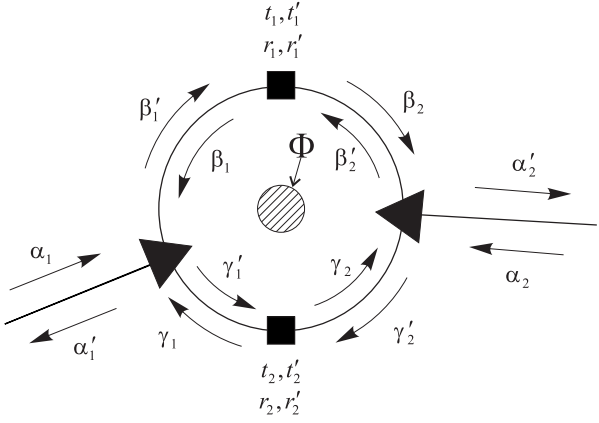}
\caption{A ring of radius $R$ connected to two leads, indicated by triangular
arrows, with two elastic scatterers indicated by the black squares.
The transmission and reflection amplitudes of the scatterers are
denoted by $t_i$ and $r_i, i=1,2$, respectively. A flux $\Phi$
pierces the ring through its center} \label{f1}
\end{figure}
\begin{equation}
\bm{\alpha}^\prime=\bm{S}\bm{\alpha}.
 \label{eq1}
\end{equation}

Current conservation implies that ${\bf S}$ is unitary and
time-reversal invariance, when applicable, entails ${\bf S}^*={\bf
S}^{-1}$. This means that the ${\bf S}$ matrix is symmetric
\cite{but}. Further, we assume that it is real. Then ${\bf S}$ is
given by
\begin{equation}
S=\left[
\begin{array}{ccc}
a & \ b & \ c \\
b & \ d & \ e\\
c & \ e & \ f
\end{array}
\right]. \label{smat}
\end{equation}

If the ${\bf S}$ matrix is symmetric with respect to both arms, one
takes $b=c$ and $d=f$,
see Fig. 1 and, e.g.,  Eqs.  (A7)-(A9).  We introduce an asymmetry by taking
$b=\lambda c$.
Then, as detailed in
appendix A, the unitarity of the ${\bf S}$ matrix leads to the
following form,
\begin{equation}
S=\left[
\begin{array}{ccc}
a & \ \
\lambda\nu & \ \ \nu\\
\ \\
\lambda\nu & \ \ \ \  \eta-a & \ \
-\lambda\eta \\
\ \\
\nu & -\lambda\eta
& \ \ 1-\eta
\end{array}
\right] \label{smat1}
\end{equation}
where $\nu=(1-a^2)^{1/2}/\mu, \
\eta=(a+1)/\mu^2, \ \mu=(\lambda^2+1)^{1/2},$ and $-1 < a < 1.$ Equation (3) is also valid for
$\lambda=1$ and the determinant of ${\bf S}$ is invariant under the
change $\lambda\to 1/\lambda$.
This change reflects the fact that
the results should be the same when the asymmetric injection favors equally
the upper or lower  arm.

The reflection probability at the left junction is $a^2$. Perfect
reflection entails $a=1$  and perfect transmission $a=0$. These two
limits correspond to those of $\epsilon$ in Ref.  \cite{but} being,
respectively, $0$ and $1/2$. It can be invoked that the probability of perfect
reflection at the left junction should be independent of the asymmetry of the ring with respect to the two arms.
Therefore, one should have  $a^2=1-2\epsilon$. It will be shown later in a different way that  indeed one has
$a^2=1-2\epsilon$.

\section{Transmission amplitude}

 Equation (1) relates the
amplitudes of the incoming waves to those of  the outgoing ones. For
the usual scattering from the left we take $\alpha_1=1$ and
$\alpha_2=0$. The corresponding transmission is given by
$T=|\alpha_2^\prime|^2$. We proceed along the lines of Ref.
\cite{but}. We write Eq. (1) as $\bm{ \alpha}_2^\prime={\bf S} \bm
{\alpha}_2$ for the right junction and as $\bm{
\alpha}_1^\prime={\bf S}\bm{ \alpha}_1$ for the left junction. The
connection between the two junctions is made by writing
\begin{equation}
\left(
\begin{array}{c}
\beta_2\\
\beta_2^\prime
\end{array}
\right)
=e^{-i\theta_1} {\bf t}_1 \left(
\begin{array}{c}
\beta_1^\prime\\
\beta_1
\end{array}
\right) \label{smat}
\end{equation}
for the amplitudes in the upper arm, where
 ${\bf t}_1 $ is the  matrix describing the transfer
 through scatterer 1, cf. Fig. 1, and $\theta_1$ a phase shift  introduced by the flux
$\Phi=\pi R^2 B$, $B$ being the magnetic field and $R$ the ring's radius. The  matrix ${\bf
t}_1 $ is given by
\begin{equation}
{\bf t}_1=\left(
\begin{array}{cc}
1/t^*\ \ &\ \ \ -r^*/t^* \\
-r/t\ \  & \ \ \ 1/t
\end{array}
\right), \label{smat}
\end{equation}
with $r$ and $t$ the reflection and transmission amplitudes,
respectively. A similar expression transfers the amplitudes in the
lower arm and involves a phase shift $\theta_2$. These shifts
satisfy the relation  $\theta_1 + \theta_2=2\pi\Phi/\Phi_0$, where
$\Phi_0=h/e$ is the flux quantum. Using these expressions and
solving the systems of equations $\bm{ \alpha}_2^\prime={\bf S}\bm{
\alpha}_2$ and $\bm{ \alpha}_1^\prime={\bf S}\bm{ \alpha}_1$, as
detailed in  appendix A, we obtain the  expression for
$\alpha_2^\prime$.

    We will consider only two cases: i) no scatterers are present in
the ring's arms, and ii) one scatterer is present in one arm. The
case with both scatterers present can be treated in the same way, 
see 
Ref.  \onlinecite{but} for $\lambda=1.$

    For case i) we have $t_1=t_2=e^{i\phi}, r_1=r_2=r_1^\prime=r_2^\prime=0$,
$\theta_1=\theta_2=\pi\Phi/\Phi_0$, and $\phi$ the phase change of
the transmitted wave. It is related to the energy $E$ by \cite{mah}
$\phi=(2m^*E)^{1/2}(\pi R/\hbar)$. Then, using Eq. (A.20) of
appendix A,  we obtain
\begin{equation}
\alpha_2^\prime=\frac{-2i(a^2-1)\Lambda
e^{-i\theta}(\lambda^2+e^{2i\theta})\ \sin\phi  }
{(a+1)^2 \Lambda_\theta 
- \Lambda^2[F(\phi,a)+2a]}, \label{amf}
\end{equation}
where $\Lambda=\lambda^2+1$, $\Lambda_\theta=\lambda^4+2\lambda^2 \cos2\theta
+1$, and
$F(\phi,a)=(a^2+1)\cos2\phi+i(a^2-1)\sin2\phi$. Then the
transmission is given by
\begin{equation}
T=\frac{4(a^2-1)^2\Lambda^2
\Lambda_\theta \ 
\sin^2\phi  } {\left\{(a+1)^2 \Lambda_\theta 
-
\Lambda^2 a_\phi 
\right\}^2+(a^2-1)^2\Lambda^4
\sin^22\phi},
 \label{amf}
\end{equation}
where 
$a_\phi=(a^2+1)\cos2\phi+2a$.

Surprisingly, for $\theta=0$ we have $ \Lambda_\theta = 
\Lambda^2$ and the dependence on $\lambda$ disappears.
Then  $\alpha_2^\prime$ takes the much simpler form
$\alpha_2^\prime=2i(a^2-1) \sin\phi/[F(\phi,a)-(a^2+1)]$. This
simplifies the  transmission considerably,  as it  takes the form
\begin{equation}
T=\frac{4(a^2-1)^2 \ \sin^2\phi  }
{\left\{(a+1)^2 - a_\phi 
\right\}^2+(a^2-1)^2
\sin^22\phi}
 \label{amf}
\end{equation}

 For case ii) we have $t_1=T_s^{1/2}e^{i\phi}, t_2=e^{i\phi}, r_1=r_1^\prime=R_s^{1/2}
e^{-i(\pi/2-\phi)}, r_2=r_2^\prime=0$,
 and $\theta_1=\theta_2=\pi\Phi/\Phi_0$. $T_s$ is the transmission amplitude
 of the   scatterer and $R_s=1-T_s$
 the reflection amplitude. Then $\alpha_2^\prime$
 takes the form
 \begin{equation}
\alpha_2^\prime=\frac{-2i(a^2-1)\Lambda
 \ e^{i\theta}[(
\lambda^2T_s^{1/2}e^{-2i\theta}+1) \sin\phi-R_s^{1/2}]}
{(a+1)^2 \Lambda_{s\theta} 
-\Lambda^2[F(\phi,a) +2R_s^{1/2}
G(\phi,a)/\Lambda +2a]},\label{amf}
\end{equation}
where  $\Lambda_{s\theta}=\lambda^4+2\lambda^2 T_s^{1/2}\cos2\theta
+1$ and $G(\phi,a)=[(a^2-2\lambda a+1)\sin\phi -i(a^2-1)\cos\phi]$. For
$R_s\to 0$ we have $T_s=1$ and Eq. (9) reduces to Eq. (6). In
contrast though with Eq. (6), the $\theta=0$ limit of Eq. (9) does
depend on $\lambda$. The corresponding expressions though for
$\alpha_2^\prime$ and $T$ are too lengthy and will not be given here.

It is interesting to combine the asymmetric injection represented
by $\lambda$ with an asymmetry in the ring's arms due to different
average densities \cite{ped}. Without any scatterer in the arms
this asymmetry can be modelled by taking $t_1=e^{i(\phi+\delta)}$
and $t_2=e^{i(\phi-\delta)}$ and the same other parameters as in
case i). The resulting expression for $\alpha_2^\prime$ can be
written as
 \begin{equation}
\alpha_2^\prime=\frac{-2i(a^2-1)\Lambda
 \ e^{-i\theta}[
\lambda^2\sin(\phi-\delta)+ e^{2i\theta}\sin(\phi+\delta)]}
{(a+1)^2 \Lambda_{\theta\delta} 
-\Lambda^2[F(\phi,a)
+2a\cos2\delta]+G(a,\delta)}, 
\end{equation}
where $\Lambda_{\theta\delta}=\lambda^4+1)\cos2\delta +
2\lambda^2\cos2\theta$ and  $G(a,\delta)=i(a^2-1)(\lambda^4-1)\sin2\delta$. For
$\delta=0$ Eq. (10) reduces to Eq. (6) while for $\lambda=1$ it
gives Eq. (3)  of Ref. \onlinecite{ped} with which an experimentally
observed period halving of the A-B oscillations was explained.
Notice further that, in contrast with Eq. (6), the dependence on
$\lambda$ does not disappear from Eq. (10) if we set $\theta=0$.

\section{Results}

We now present numerical results for the transmission $T$ given by
$T=|\alpha_2^\prime|^2$. We evaluate $T$ using Eq. (6) and plot it
vs $\phi$ in   Fig. 2(a)  for two different values of the
flux. 
In   Fig. 2(b) we plot $T$ as a
function of the flux $\Phi/\Phi_0$ for two different values of
$\phi$. In both panels we have $a=0$. This value of $a$ corresponds
to $\epsilon=1/2$ in Ref.  \cite{but} and the figure is intending to
show the effect of the {\it asymmetric}  injection ( $\lambda\neq 1$) with respect to both ring
arms: $\lambda=1$ pertains to solid and dashed curves and $\lambda=3$ to
 dotted and dash-dotted
curves. For $\lambda=1$ the results coincide with those of Ref.
\onlinecite{but}.

\begin{figure}[tpb]
\includegraphics*[width=75mm]{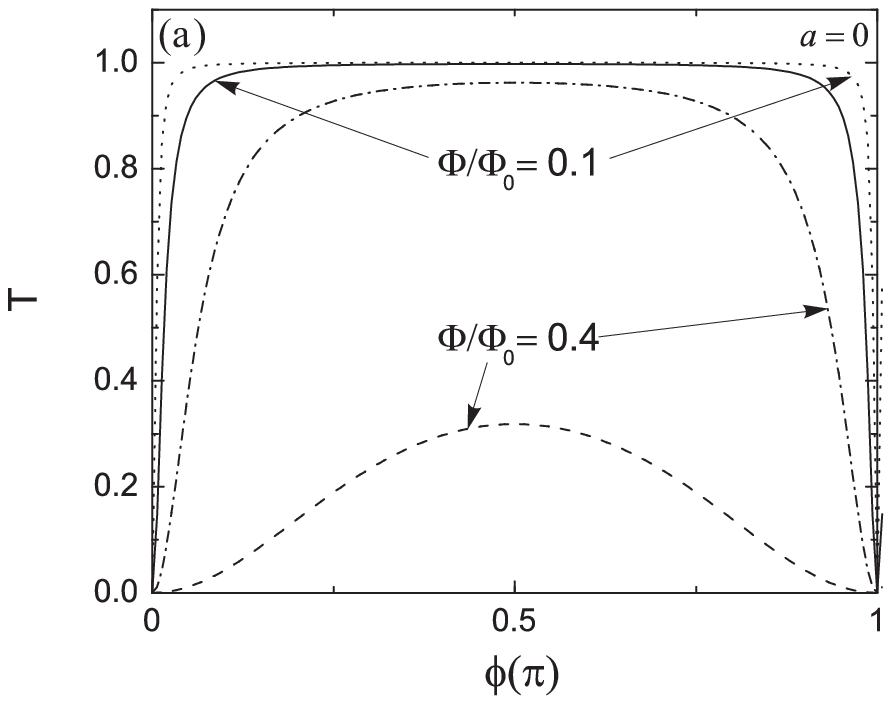} \\
\includegraphics*[width=75mm]{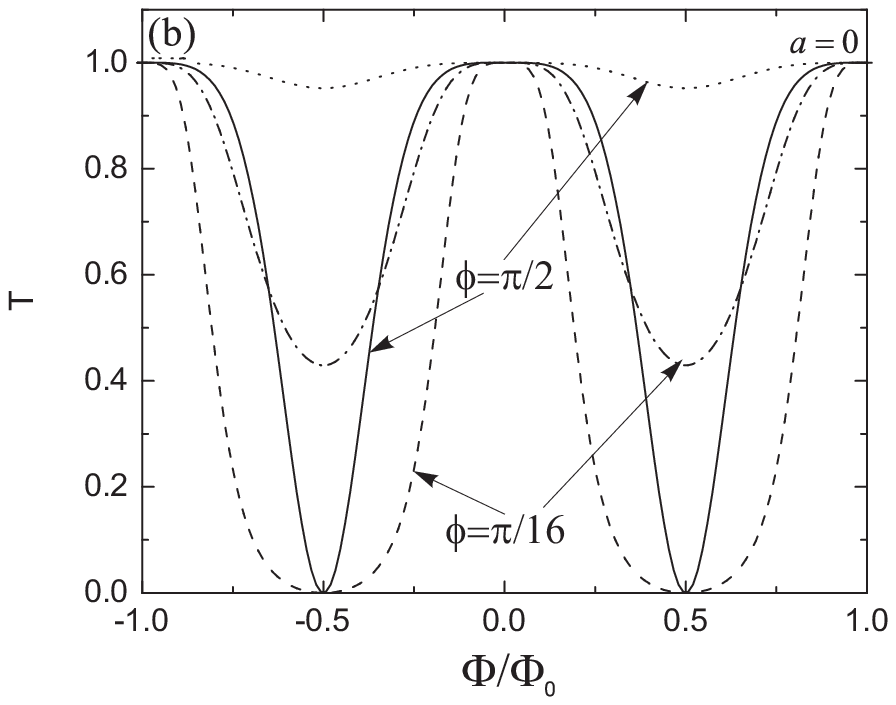} \caption{ Transmission vs $\phi$ (a) for two values of the flux and
vs flux (b) for two values of $\phi$. The solid and dashed curves
are for $\lambda=1$ and the dotted and dash-dotted curves for
$\lambda=3$. The value of $a$ is zero, i.e., there is no reflection at the left
lead-ring junction.} \label{f1}
\end{figure}

If a scatterer is present in one of the arms, we evaluate $T$ using
Eq. (9) and plot it in Fig. 3 for $T_{s}=0.25$. Panels and curves are marked as in
Fig. 2. As can
be seen, a major difference between the two cases is the phase shift
introduced by the scatterer in the $T$ vs $\Phi/\Phi_0$ panel and
the asymmetry between the left and right parts in the $T$ vs $\phi$
panel. The sinusoidal dependence of the transmission on  $\phi$ or
the flux $\Phi/\Phi_0$ stems directly from that of the transmission,
cf. Eq. (7), and especially from that of the numerator for
$\lambda>1$.

\begin{figure}[tpb]
\includegraphics*[width=75mm]{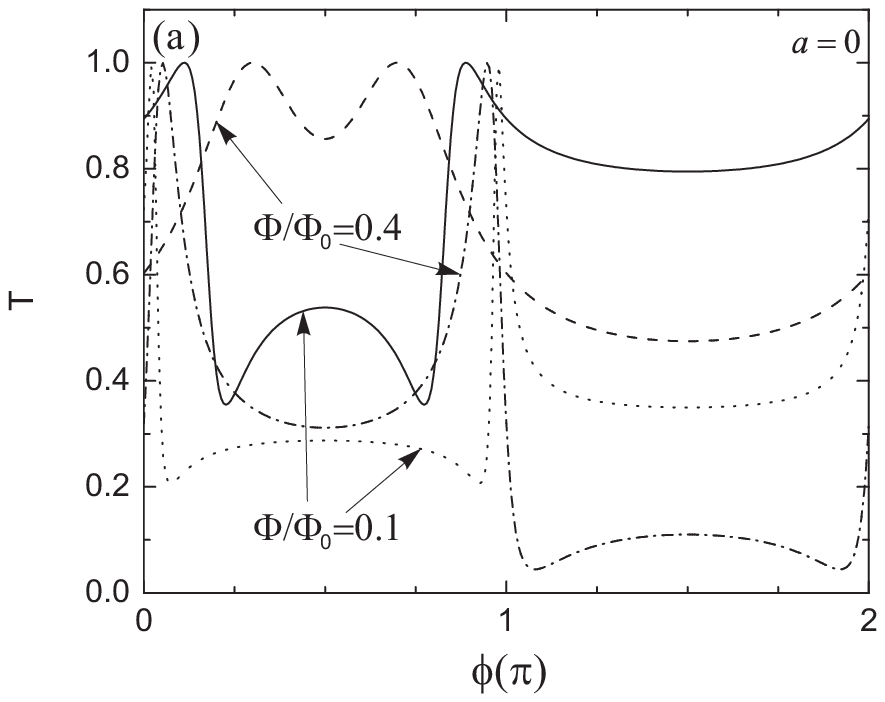} \\
\includegraphics*[width=75mm]{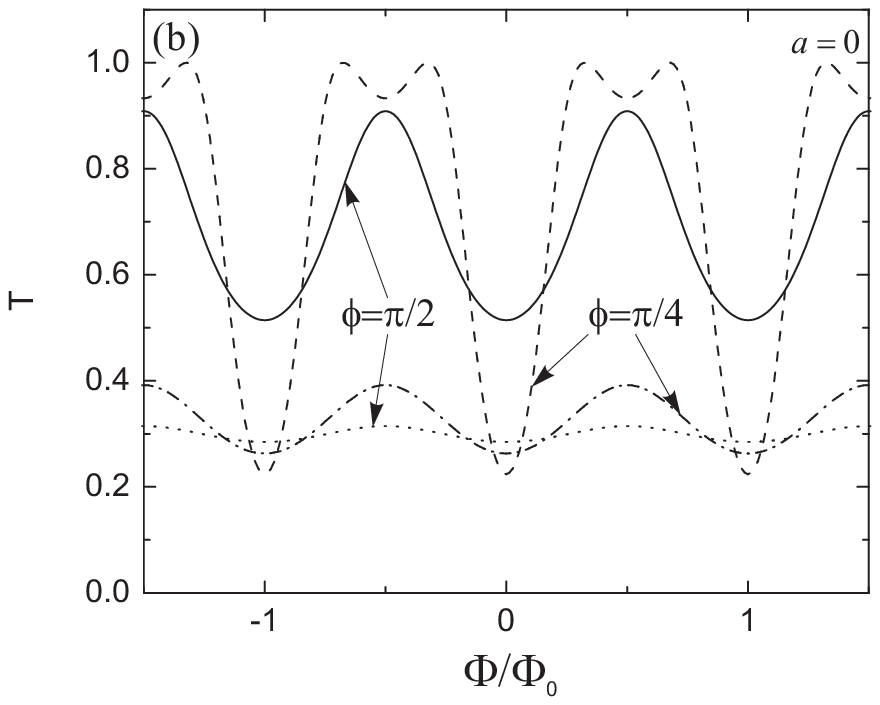}
\caption{Transmission vs $\phi$ (a) and vs flux (b) when one
scatterer is present in one of the arms with strength $T_{s}=0.25$. The curves are marked as in
Fig. 2.} \label{f1}
\end{figure}

We now consider only the asymmetric case, fix the value of
$\lambda$ ($\lambda=3$), and focus attention on the dependence of
the transmission on the parameter $a$. We show the results in Fig.
4 only for case i), i.e., when no scatterer is present. When one is
present,  a phase shift of $\pi/2$ occurs in the T versus $\phi$
curves and maxima are converted into minima and {\it vice versa}.
The flux is set to $\Phi/\Phi_0=0.4$ in panel (a)  and $\phi$ is
set to $\pi/2$ in
 panel (b). As expected from Eqs. (7) and (9), the transmission
$T$ decreases  with increasing $a$ when $\phi$ is fixed.

\begin{figure}[tpb]
\vspace{0.5cm}
\includegraphics*[width=75mm]{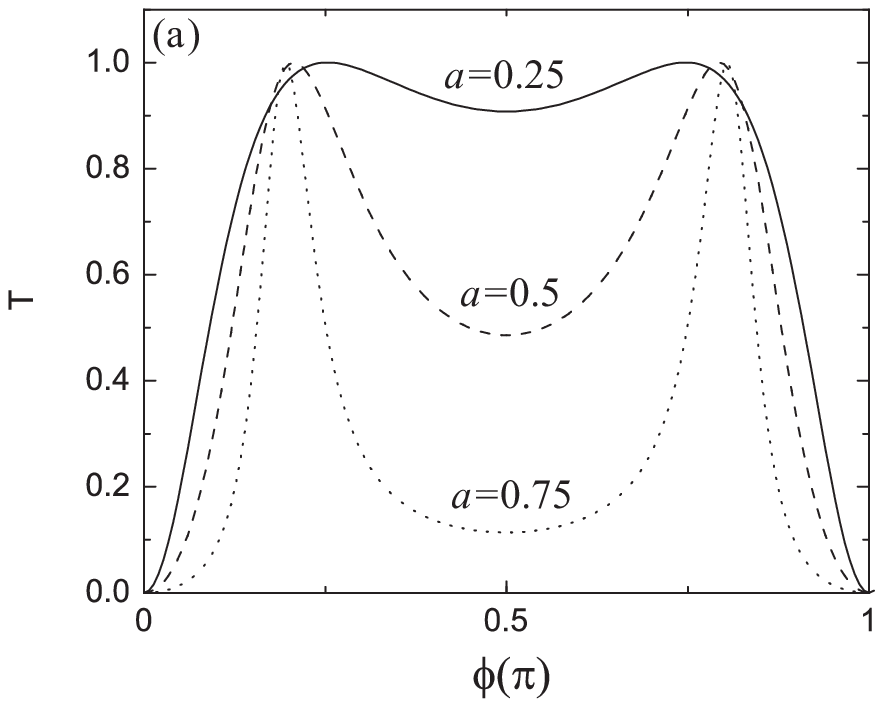} 
\includegraphics*[width=75mm]{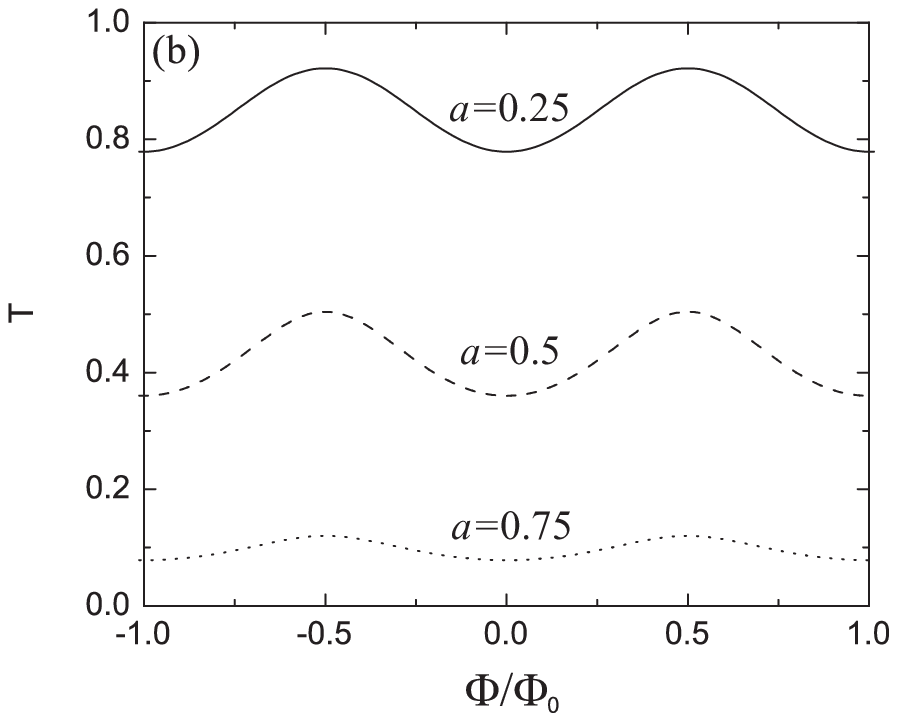} 
\caption{Transmission vs $\phi$ (a) for $\Phi/\Phi_0=0.4$ and vs flux (b) for $\lambda=3$
and $\phi=\pi/2$ with  $a$ as indicated.} \label{f1}
\end{figure}

The results  of Figs. 1-4 show that the parameter $\lambda$,
which has to be real for the S matrix to be unitary,
affects the amplitude of the transmission but not its phase.
We compared the conductance $G$ resulting from Eq. (10) with the experimental
 results of the right panel of Fig. 3 of Ref.
12. We used $\lambda=1$, $\alpha=0$ corresponding to $\epsilon=1/2$, and the relation of $\delta$ with
$k_F$ given in this work. Though the agreement is a bit better than that reported in Ref. 12,
especially with regard to the oscillation amplitude of $G$ vs $\phi$, determined by the Fermi level, or $G$ increase, it remains qualitatively the same and changing $\lambda=1$ to $\lambda\neq 1$, with $a=0$ or
$a\neq 0$, brings only a  minor quantitative improvement.

\section{Comparison with other approaches, rings of finite width}

The results presented so far are valid for rings of zero width or
rings whose width is much smaller than their radii and only the
lowest, in the radial direction, energy level is occupied. The
question then arises i) how the results compare to those of other
approaches, and ii) how  relevant they are to rings of finite width.

With regard to point i) one frequently followed approach is to
consider the same 1D geometry but employ Griffith's boundary
conditions at the junctions between the leads and the ring
\cite{cho}. Neither the {\bf S} matrix approach  nor
these conditions apply to rings of finite width. One then has to
resort to either  pure and often heavily involved numerical
calculations that pertain to point ii) \cite{fru,sza} or
an appropriate modification of  Griffith's  conditions \cite{voo}.
We first address point i) and then point ii).

Comparing the zero magnetic field limit of Eq. (24) of Ref.
\onlinecite{but} for the transmission,
$T=|\alpha_2^\prime|^2=1/[(1-2\epsilon)\sin^2\phi/\epsilon^2+1]$,
with the zero SOI limit of $T$  in Ref. \onlinecite{cho}, $T=1/[9
\sin^2\phi/16+1]$,  obtained by applying Griffith's boundary
conditions, one finds that the results coincide when the parameter
$\epsilon$ is equal to $4/9$ which is close to the upper limit
$\epsilon=1/2$. If we compare any of these expressions with ours, as
obtained using Eq. (8),  for any $\lambda$ we obtain $a=\pm
(1-2\epsilon)^{1/2}$ in the first case and $a=\pm 1/3$ in the
second.

Surprisingly, the same value of $\epsilon$ is obtained for a non
zero magnetic field if one disregards the fact that
$\theta$ in  Eq. (24) of Ref. \onlinecite{but} appears as $\theta/2$ in
Ref. \onlinecite{cho} since the former is given by
$T=4\sin^2\phi\cos^2\theta/[(2a^2/\epsilon)\sin^2\phi+(b^2/\epsilon)
(\cos2\theta-\cos2\phi]^2, a^2+b^2=1-\epsilon$, and the latter by
$T=4\sin^2\phi\cos^2(\theta/2)/[\sin^2\phi/2+
(\cos\theta-\cos2\phi]^2$. Unfortunately, attempting to make the
same comparison between any of these expressions for $T$ and ours,
using Eq. (7) with $\lambda\neq \pm 1$, gives a very unwieldy result
that involves transcendental equations.

One way to proceed with case ii), i.e., with rings of finite width,
is to slightly modify
Griffith's conditions so that the width appears in them and in the
 expression for the transmission. This is done \cite{voo} at the expense of
 an additional
parameter $\nu$, of order 1, whose value is obtained from  a
comparison of the transmission with an exact calculation. We have
done so for the transmission through a ring of width $W$ at zero
magnetic field and give the result in Appendix B, cf. Eq. (B2).
Equating this result to that of Ref. \onlinecite{but} gives
$\epsilon=4/(\mu+9)$ with $\mu=4\nu^2/k^2W^2$, while equating it to
ours leads to a value of $a$ determined from $(a^2+1)/(a^2-1)=-
[(\mu+1)(\mu+9)+1]^{1/2}$.
Notice that this determination of $\epsilon$ and $a$ makes them
depend on  the energy, through the wave vector $k$, and the width $W$. We assume that
approximately the same values are obtained for $B\neq 0$.

A second way to proceed is to compare directly  our result for
$W=0$ with an exact numerical one and try to fit the latter by
varying the parameter $\lambda$. We have done so with
$\lambda=c/(1+(\Phi/\Phi_0)^s)$ in order to mimick the asymmetry
reported in Ref. \onlinecite{sza} and attributed to the effect of the
Lorentz force in a quantum wire of finite width. In the left panel
of Fig. 5 we show the transmission as a function of the flux. The
values of $\phi$ used in producing the solid, dashed, and dotted
curve correspond to the wave vectors $k=0.091$/nm, $k=0.06$/nm,
and $k=0.053$/nm in Fig. 8 of Ref. \onlinecite{sza}. The values of  $a,
s$ and $c$ used are $a=0.25, s=1.25$ and $c=1$. The qualitative
agreement between the two results is very good with regard to the
period of the oscillations and the height of the transmission
peaks which in our case are more rounded than in Ref. \onlinecite{sza}.
Notice in particular the reduction of the transmission minima with
increasing $\Phi/\Phi_0$ which is attributed to the influence of
the Lorentz force and is well reproduced. Of course our simple
model cannot reproduce all the details of Ref. \onlinecite{sza} but its
analytical simplicity is an advantage over the approach of Ref.
\onlinecite{sza}.

The asymmetry reported in Ref. \onlinecite{sza} 
is also exhibited in the right panel of Fig. 5 where we plot $T$
versus $\Phi/\Phi_0$ for case ii), i.e., when one scatterer is
present in one arm. We plot $T$ for different $T_s$ using
$\phi=6.95\pi$. The dashed-dotted, dotted, dashed, and solid
curves correspond to $T_s=0.25, 0.5, 0.75$, and $1$, respectively.
We also used $a=0.4, s=1.25$ and $c=1$. The results are similar to
those shown
 in the left panel, where different wave vectors were
used from one curve to another. As can be seen, the overall trend
is similar to that of Fig. 12 of Ref. \onlinecite{sza}. A detailed
comparison cannot be made though because here we consider
transmission through a barrier whereas Ref. \onlinecite{sza} considered
it over a 
well. Furthermore, in Ref. \onlinecite{sza} the transmission of a Gaussian wave packet, with a spread in energy, was
 investigated, which can be deconvoluted into a series of plane waves
 around an average wave number, while here 
 the
 transmission of a plane wave with a well-defined energy or wave vector was studied.
In principle we could take the injected, reflected, and
transmitted wave packages of Ref. 6, Fourier transform them, and
obtain the transmission and reflection coefficients for each wave
vector, i.e., for each energy, and add this difference to the
manuscript. However, the aim here is not to present an exact comparison
between these two approaches, which is unrealistic because of the
different models used, but to show that we can simulate the
Lorentz-force-induced asymmetric injection of  electrons into the
arms of the rings with a simple analytical calculation. 
 We expect though that such a calculation will lead to some minor quantitative discrepancies in the transmission even if the  approach of Ref. 6 is applied to a barrier, as in our case,
due to the spread in the energy of the incident electrons. 

\begin{figure}[tpb]
\includegraphics*[width=75mm]{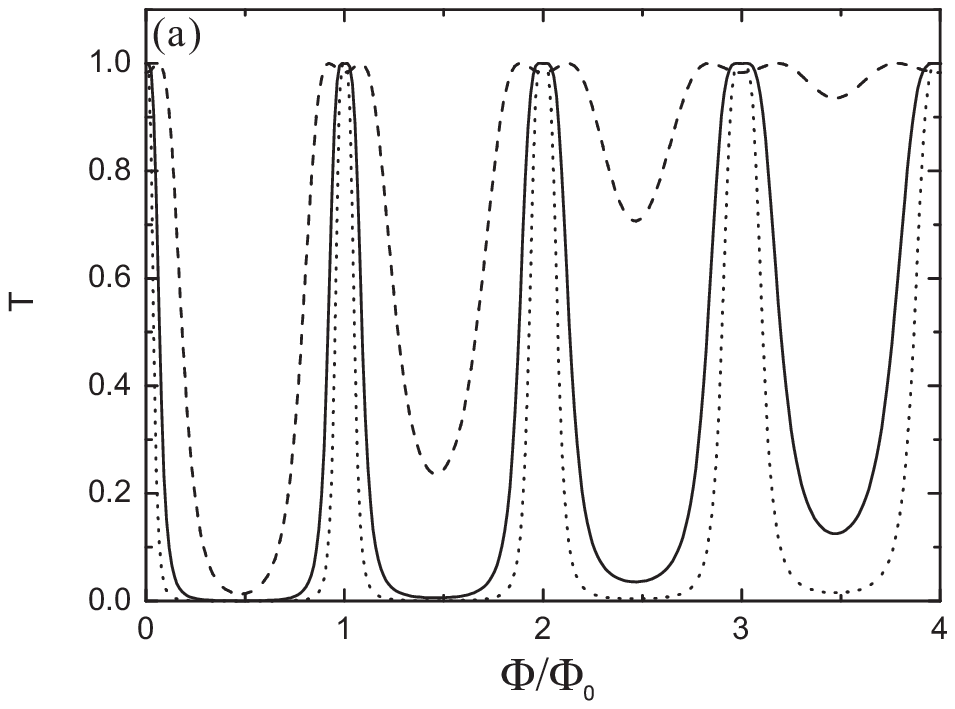} \\
\includegraphics*[width=75mm]{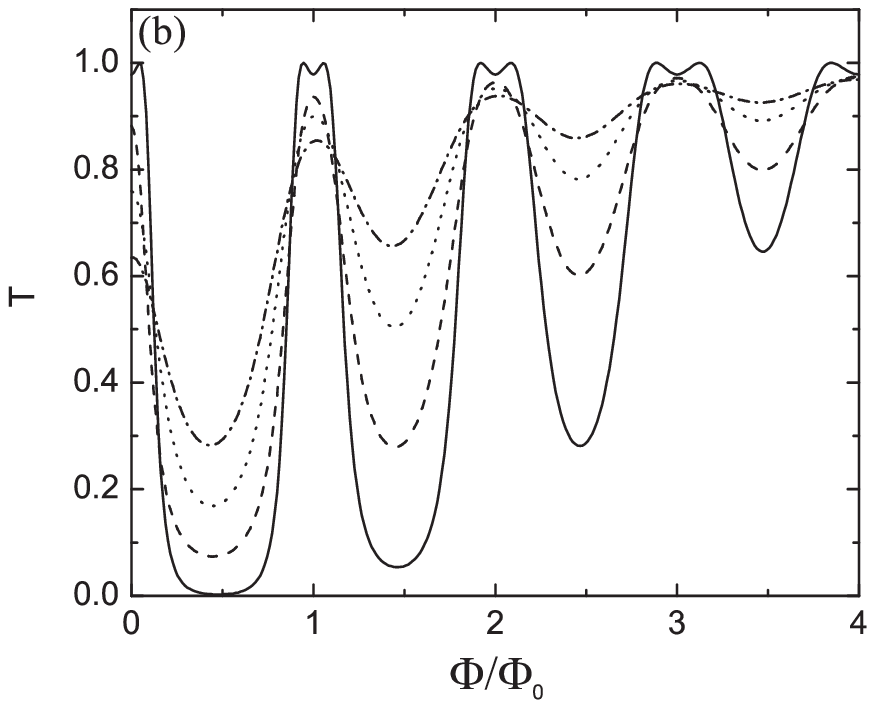}
\caption{Left panel. Transmission vs flux with $\phi$ such that
the solid, dashed, and dotted curve correspond to the wave vectors
$k=0.091$/nm, $k=0.06$/nm, and $k=0.053$/nm in Fig. 8 of Ref.
\onlinecite{sza}. The values of  $a$, $s$ and $c$ used are $a=0.25$,
$s=1.25$ and $c=1$. No scatterer is present in the ring's arms.
Right panel. As in left panel with one scatterer  present in one
of the arms. The dashed-dotted, dotted, dashed, and solid curves
are for 
$T_s=0.25, 0.5, 0.75$, and $1$, respectively. The
other parameters are $\phi=6.95\pi$, $a=0.4$, $s=1.5$ and $c=1$. 
}
\label{f6}
\end{figure}


\section{Concluding remarks}

We evaluated the transmission through a mesoscopic ring, in the
presence of a AB flux, using an ${\bf S}$ matrix that is not
symmetric with respect to the two arms of the ring. All elements
were expressed in terms of two parameters, $a$ and $\lambda$, the
latter expressing the asymmetry through $\lambda=b/c$, with $b, c$
the pertinent elements of the ${\bf S}$ matrix. 
The
determinant of the ${\bf S}$ matrix  and the transmission are
invariant under the change $\lambda\to 1/\lambda.$ Previous
results of the literature pertaining to a {\it symmetric} ${\bf
S}$ matrix were readily recovered for $\lambda=1$. The dependence
on  the parameter
 $\lambda$ disappears from the transmission, see
Eqs. (6) and (7), but not from the more general result given by
Eq. (10) when the flux is zero. It is important mostly for a non
zero flux and, depending on the parameters, it modifies the
results considerably as shown in Figs. 2 and 3 where results for 
$\lambda=1$ and $\lambda=3$ are
contrasted.

We also evaluated the transmission when an asymmetry was
introduced externally, i.e., when one scatterer was placed in one
of the ring's arms. Of course this applies to both cases
$\lambda=1$ and $\lambda\neq 1$. The transmission shows a rich
structure as a function of the parameter $a$, $\phi$ (or wave
vector), $\Phi/\Phi_0$, and the strength $T_s$, cf. Figs. 1-5. The
results for different $a$ shown in Figs. 4, 5 correspond to those
for $\epsilon$ in Ref. \onlinecite{but} in which no results as a
function of the flux or the strength $T_s$ were shown.

Importantly, results on rings of finite width could be mimicked in
two ways. In one way we simply fitted the results of a heavy
numerical treatment \cite{sza} by using
$\lambda=c/[1+(\Phi/\Phi_0)^s],$ with $s$ between $1$ and $2$. The
two results are in good qualitative agreement especially in the
case when no scatterer was present in the arms, cf. Fig. 5. When a
scatterer was present a real comparison could not be made due to
the different nature of the scatterers involved, a barrier in our
case, a Gaussian well in that of Ref. \onlinecite{sza}. Since the
stronger the magnetic field the stronger the asymmetry obtained in
Ref. \onlinecite{sza}, it's natural to expect that it should be
reflected in $\lambda$ even if our treatment applies only to rings
whose width is much smaller than their radii and only the lowest,
in the radial direction, energy level is occupied.  As we saw the
overall trend in $T$ vs $\Phi/\Phi_0$ was reproduced quite well
though not all the details. We emphasize that this agreement
cannot be obtained with an $S$ matrix that is {\it symmetric} with
respect to both arms, i.e., for $\lambda=1$.

Despite the agreement just mentioned above, the model has its
limitations as was pointed out at the end of Sec. IV, regarding a
further comparison between its results and those  of Ref.
\onlinecite{sza}, and at the end of Sec. III regarding a possible
improvement of the agreement, upon using Eq. (10) with
$\lambda\neq 1$, between the reported experimental \cite{ped} and
theoretical results. Also, if  the potentials in the two arms are
different, complex interference patterns may result and the
asymmetric injection, affecting only the amplitude of the
transmission, is unlikely to completely describe the systems
discussed in this paper.

In another way we used results of the literature that incorporate the width of the ring in the boundary
conditions, though not rigorously, to determine the parameters $a$ and $\epsilon$ by comparing  results for the
transmission between different approaches. This made $a$ and $\epsilon$ depend on the energy and the width of
the ring. In addition,  through this comparison we demonstrated
the equality $a^2=1-\epsilon$.

Another externally imposed asymmetry that could be considered is to have the lengths of the two arms unequal.
This has been treated in Ref. \onlinecite{col} using a {\it symmetric} ${\bf S}$ matrix and, as expected, lead to a certain dephasing of the AB oscillations.
A similar study, involving
arms of unequal length and a $4 \times 4$ ${\bf S}$ matrix  in the presence of SOI but without
scatterers in the arms, appeared recently \cite{aeb}.

A possible extension of the theory presented here would be to consider in  detail the real nature of the scatterers, barriers or wells,
placed in one arm and have an explicit  energy dependence in the transmission ($T_s$) and reflection ($R_s)$ probabilities  instead of taking them as parameters as we did.
  Another extension would be to consider a chain of rings with periodic modulations in the magnetic field   or ring radius in analogy with a recent work on rings in the presence of SOI  \cite{mol2}. 
  A last extension concerns the determination $\lambda$. Since $\lambda$, like $\epsilon$  or $a$, is a  parameter constrained only by the unitarity of the ${\bf S}$ matrix (it has to be real), it cannot  be evaluated. The only way it could be  determined is to compare two transmission results obtained by two different methods as we have done in Sec. V for the parameters $\epsilon$ and $a$. Presently  such results are not available.  
  All these extensions are left for future work. We expect  that our results,
  though incomplete in some respects, will be tested by appropriate experiments.
\ \\


 This work was supported by the Flemish-Hungarian bilateral program, the Canadian NSERC Grant No.
OGP0121756,   the Flemish Science Foundation (FWO-VI),
and the SANDiE EU-network of excellence. O. K. is supported by the Marie Curie training project. O. K.  and  M. G. B. are also supported by the  Hungarian Scientific Research Fund
under Contract No. 48888.
\vspace*{-0.2cm}
\appendix
\section{}
In part i)  below we determine the elements of the ${\bf S}$ matrix
and in part ii) we solve the systems of  equations $\bm{
\alpha}_2^\prime={\bf S}\bm{ \alpha}_2$ and $\bm{
\alpha}_1^\prime={\bf S}\bm{ \alpha}_1$.

i) The unitarity of the ${\bf S}$ matrix, cf. 
Eq. (2), leads to
the following 
relations between   its elements $a, b, c, d, e,
f$.
\begin{eqnarray}
a^2+b^2+c^2=
b^2+d^2+e^2=
c^2+e^2+f^2=1\\*
ab+bd+ce=
ac+be+cf=
bc+de+ef=0
 \label{eq4}
\end{eqnarray}
To reduce the number of parameters and make the two arms not
equivalent to each other, we take $b=\lambda c$.
Then Eqs. (A2) give $e=-\lambda (a+d)$ and $f=-a-d\pm 1.$
With $b=\lambda c$ Eqs. (A1) give $b=\pm\lambda(1-a^2)^{1/2}/\mu$
where $\mu=(\lambda^2+1)^{1/2}$. Obviously $1-a^2\geq 0$, i.e.,
$|a|\leq 1.$ Then Eqs. (A1) and $e=-\lambda (a+d)$ determine $d$ as
$d=(\lambda^2a-1)/\mu^2$. Thus, all elements can be expressed in
terms of $a$ and $\lambda$ and the result is given by Eq. (3).  Reference
\onlinecite{but} took $\lambda=1$ and instead of $a$ used the parameter
$\epsilon$, $0\leq \epsilon \leq 1/2$. The value $\epsilon =1/2$ corresponds
to $a=0$ and $\epsilon =0$ to $a=1$.

 ii) For the right junction we obtain
\begin{eqnarray} \alpha_2^\prime&=&b\beta_2+c\gamma_2\\*
 \beta_2^\prime
&=&d\beta_2+e\gamma_2\\* \gamma_2^\prime &=& e\beta_2+f\gamma_2.
 \label{eq1}
\end{eqnarray}
Using Eqs. (A4)-(A5) we can write
\begin{equation}
\left(
\begin{array}{c}
\gamma_2^\prime\\
\gamma_2
\end{array}
\right)={\bf t}_{l2} \left(
\begin{array}{c}
\beta_2\\
\beta_2^\prime
\end{array}
\right) \label{smat}
\end{equation}
where ${\bf t}_{l2}$ is the matrix
\begin{equation}
{\bf t}_{l2}=\frac{1}{e}\left(
\begin{array}{cc}
e^2-fd\ \ &\ \ \ f \\
-d\ \  & \ \ \ 1
\end{array}
\right) \label{smat}
\end{equation}
Notice that $det({\bf t}_{l2})=1$.

 For the left junction we obtain
\begin{eqnarray}
\alpha_1^\prime&=& a+b\beta_1+c\gamma_1\\*
 \beta_1^\prime &=&
b+d\beta_1+e\gamma_1\\*
\gamma_1^\prime &=& c+e\beta_1+f\gamma_1
 \label{eq4}
\end{eqnarray}
Using Eqs. (A9), (A10) we can write
\begin{equation}
\left(
\begin{array}{c}
\beta_1^\prime\\
\beta_1
\end{array}
\right)=\frac{b}{e} \left(
\begin{array}{c}
be-dc \\
-c
\end{array}
\right)+ {\bf t}_{l1} \left(
\begin{array}{c}
\gamma_1\\
\gamma_1\prime
\end{array}
\right),\label{smat}
\end{equation}
where ${\bf t}_{l1}$ is given by ${\bf t}_{l2}$ with $d$ and $f$
interchanged.

The connection between the two junctions is made 
by writing
\begin{equation}
\left(
\begin{array}{c}
\beta_2\\
\beta_2^\prime
\end{array}
\right) =e^{-i\theta_1} {\bf t}_1 \left(
\begin{array}{c}
\beta_1^\prime\\
\beta_1
\end{array}
\right), \label{smat}
\end{equation}
for the amplitudes in the upper arm, and
\begin{equation}
\left(
\begin{array}{c}
\gamma_1\\
\gamma_1^\prime
\end{array}
\right)=e^{-i\theta_2} {\bf t}_2^\prime \left(
\begin{array}{c}
\gamma_2^\prime\\
\gamma_2
\end{array}
\right) \label{smat}
\end{equation}
for the amplitudes in the lower arm, where ${\bf t}_1$ and ${\bf
t}_2$ are the matrices associated with the first and second
scatterer given by Eq. (A7).  Combining Eq. (A6) and Eqs.
(A11)-(A13) we can write
\begin{equation}
{\bf P}\left(
\begin{array}{c}
\beta_1^\prime\\
\beta_1
\end{array}
\right)= -\frac{b}{e} \left(
\begin{array}{c}
be-cd\\
-c
\end{array}
\right) \label{smat}
\end{equation}
with
\begin{equation}
{\bf P}={\bf t}_{l1} e^{-i\theta_2} {\bf t}_2^\prime {\bf t}_{l2}
e^{-i\theta_1} {\bf t}_1 -{\bf 1} \label{smat}
\end{equation}
where ${\bf 1}$ is the unit matrix. The transmitted amplitude
$\alpha_2^\prime$ is obtained from Eqs. (A3)-(A4) ($\Lambda=\lambda^2+1$)
\begin{equation}
\alpha_2^\prime=
[(1-a^2)\Lambda]^{1/2}(\beta_2
-
\beta_2^\prime)
/\lambda 
(a+1)
 \label{smat}
\end{equation}
and the coefficients $\beta_2$ and $\beta_2^\prime$ from Eq. (A9)
after solving Eq. (A11) for 
$\beta_1$ and
$\beta_1^\prime$. Inserting these values 
of $\beta_2$ and
$\beta_2^\prime$ in Eq. (A16) gives Eq. (6) of Sec. III.

\section{}

Griffith's boundary conditions applied at a junction between the
leads and the ring, e.g., at the black triangles in Fig. 1, are i)
the continuity of the wave function and ii) the continuity of the
flux. For a ring of finite width one modifies condition ii) by
adding a term $2\nu\psi/W$ 
to the left side \cite{voo} so that it reads
\begin{equation}
\sum_{i=1}^{N}\frac{
\partial\psi}{
\partial
x_i}+\frac{
2\nu\psi}{
W}
=0,
 \label{smat}
\end{equation}
where $N$ is the number of the legs at a junction ($N=3$ here) and
$\nu$  a parameter of order $1$ to be determined from a comparison
with an exact numerical result.

We have applied these conditions to a ring of finite width connected
to two leads of the same width. On each line segment the wave
function is given by $\psi=A_i\exp^{ikx_i}+B_i\exp^{-ikx_i}$. The
 inner and outer radii of such a ring are $R-W/2$ and $R+W/2$, respectively.
 Assuming nothing is incident from the right of the right junction,
and setting $\mu=4\nu^2/k^2W^2$ we obtain  the transmission, not
given in Ref.  \onlinecite{voo},   as
\begin{equation}
T= 
16/[(\mu+1)(\mu+9)sin^2(\pi kR)+16].
 \label{smat}
\end{equation}


\begin{references}
\bibitem{aha} Y. Aharonov and D. Bohm, Phys. Rev. {\bf 115}, 485 (1959).

\bibitem{aha1} Y. Aharonov and A. Casher, Phys. Rev. Lett. {\bf 53}, 319 (1984).

\bibitem{ber}M. V. Berry, Proc. R. Soc. London, Ser. A {\bf 392}, 45 (1984).

\bibitem{fru} Y.-S. Yi, T.-Z. Qian, and Z.-B. Su, Phys. Rev. B {\bf 55}, 10631 (1997);
D. Frustaglia and K. Richter, {\it ibid} {\bf 69}, 235310 (2004); S.
Souma and B. K. Nikolic, {\it ibid} {\bf 70}, 195346 (2004).

\bibitem{cho} T. Choi, S. Y.
Cho, C. M. Ryu, and C. K. Kim, Phys. Rev. B \textbf{56}, 4825
(1997); B. Moln\'{a}r, F. M. Peeters, and P. Vasilopoulos, {\it
ibid} \textbf{69}, 155335 (2004); X. F. Wang and P. Vasilopoulos,
{\it ibid} {\bf 72}, 165336 (2005).

\bibitem{sza} B. Szafran and F. M. Peeters, Phys. Rev. B {\bf 72}, 165301 (2005);
Europhys. Lett. {\bf 70}, 810 (2005).

\bibitem{col}  C.  Benjamin and A. M. Jayannavar, Phys. Rev. B {\bf 65}, 153309 (2002);
S. Bandopadhyay, P. S. Deo, and A. M. Jayannavar, {\it ibid} {\bf 70}, 075315 (2004);
 M. B\"{u}ttiker, SQUID '85, {\it Superconducting Quantum Interference Devices and their Applications}, edited by H. D. Hahlbohm and H. L\"ubbig (Walter de Gruyter, berlin, 1985), p. 529.

\bibitem{ya} A. Yacoby, M. Heiblum, D. Mahalu, and H.  Shtrikman, Phys. Rev.  Lett. {\bf 74}, 4047 (1995);
G. Cernicchiaro, T. Martin, K. hasselbach, D. Mailly, and A. Benoit, {\it ibid} {\bf 79}, 273 (1997);
R. Schuster, E. Buks, M. Heiblum, D. Mahalu, V. Umansky, and H.  Shtrikman, Nature (London) {\bf 385}, 417 (1997); S. Pedersen, A. E. Hansen, A. Kristensen, C. B. Sorensen, and P. E. Lindelof, Phys. Rev. B {\bf 61}, 5457 (2000);
M. Koenig, A. Tschetschetkin, E. M. Hankiewicz, J. Sinova, V. Hock, V. Daumer, M. Schaefer, C. R. Becker, H. Buhmann, and L. W.  Molenkamp, cond-mat/0508396; Y.  Ji, Y.  Chung, D. Sprinzak, M. Heiblum, D. Mahalu, and H.  Shtrikman, cond-mat/0303553;

\bibitem{but} M. B\"{u}ttiker, Y.
Imry, and M. Ya. Azbel, Phys. Rev. A {\bf 30}, 1982 (1984).

\bibitem{gef}Y. Gefen, Y. Imry, and M. Ya. Azbel, Phys. Rev. Lett. {\bf
52}, 129 (1984).

\bibitem{mah} We follow the usual approach of Refs. 6-7, 9-10, in which $\phi (E)$ is an independent parameter  with $E$ the Fermi level. In a more refined approach the Fermi-level of the ring depends on whether the number of the particles is even or odd, see C. H. Wu and  G. Mahler, Phys. Rev. B {\bf 43}, 5012 (1991).

\bibitem{ped} S. Pedersen, A. E. Hansen, A. Kristensen, C. B. Sorensen, and P. E. Lindelof,
  PRB {\bf 61}, 5457 (2000).

\bibitem{voo} K-K. Voo, S-C. Chen, C-S. Tang, and C-S. Chu,
 Phys. Rev. B {\bf 73},  035307 (2006).

\bibitem{aeb}  U. Aeberhard, K. Wakabayashi, and M. Sigrist,
Phys. Rev. B {\bf 72},  075328 (2005).

\bibitem{mol2} B. Moln\'{a}r, P. Vasilopoulos, and F. M. Peeters,
Phys. Rev. B {\bf 72},  075330 (2005).


\end{references}
\end{document}